\documentclass{optica-article}

\journal{opticajournal} 

\articletype{Research Article}

\articletype{Research Article}

\usepackage{bm}
\usepackage{xspace}
\usepackage{amsmath}
\usepackage{graphicx}
\usepackage{xcolor}
\usepackage{comment}
\usepackage{multirow}
\usepackage{rotating}
\usepackage{array}
\usepackage{tabularx}

\newcommand{\etal}{\textit{et al.\@}\xspace}
\newcommand{\exvivo}{\textit{ex vivo}\xspace}

\newcommand{\invivo}{\textit{in vivo}\xspace}

\newcommand{\invitro}{\textit{in vitro}\xspace}

\newcommand{\um}{\(\muup\)m\xspace}

\newcommand{\cbum}{\(\muup\)m$^3$\xspace}

\newcommand{\ift}[1]{\mathcal{F}^{-1}\left[{#1}\right]\xspace}

\graphicspath{{./figures}{.}{./figures_tmp}}

\begin{document}

\title{Dynamic-OCT simulation framework based on mathematical models of intratissue dynamics, image formation, and measurement noise}

\author{%
	Yuanke Feng,\authormark{1}
	Shumpei Fujimura,\authormark{1}
	Yiheng Lim,\authormark{1}
	Thitiya Seesan,\authormark{1}
	Rion Morishita,\authormark{1}
	Ibrahim Abd El-Sadek,\authormark{1,2}
	Pradipta Mukherjee,\authormark{1,3},
	Shuichi Makita,\authormark{1}, and
	Yoshiaki Yasuno\authormark{1,*}
}

\address{\authormark{1}Computational Optics Group, University of Tsukuba, Tsukuba, Ibaraki, Japan\\
\authormark{2}Department of Physics, Faculty of Science, Damietta University, New Damietta City, Damietta, Egypt\\
\authormark{3}Centre for Biomedical Engineering, Indian Institute of Technology Delhi, New Delhi, India}

\email{\authormark{*}yoshiaki.yasuno@cog-labs.org} 

\begin{abstract*} 
Dynamic optical coherence tomography (DOCT) enables label-free functional imaging by capturing temporal OCT signal variations caused by intracellular and intratissue motions. However, the relationship between DOCT signals and the sample motion behind them remains unclear.
This paper presents a comprehensive DOCT simulation framework that incorporates mathematical models of intracellular/intratissue motions, two OCT signal generator types that generate OCT signal time sequences from the moving scatterer models, and representative DOCT algorithms.
The theory and algorithms of the framework are described in detail, and the utility of this framework is demonstrated through numerical studies.
This framework is available as open source and will enhance the understanding and utility of DOCT.

\end{abstract*}

\section{Introduction}
Dynamic optical coherence tomography (DOCT) is a label-free imaging method that is used to visualize tissue functions \cite{Ren2024}.
DOCT has been applied to the imaging of a wide spectrum of tissues, including \exvivo animal tissues \cite{Mukherjee2021}, \invitro spheroids \cite{AbdElSadek2020,ElSadek2021} and organoids \cite{Morishita2023}, and \invivo human skin\cite{Guo2024}. 
To obtain DOCT image contrast, multiple OCT frames are captured at the same location on a sample as a time sequence.
Several DOCT signal types can be computed from such time sequences, including types that are sensitive to the magnitude of the signal fluctuations \cite{AbdElSadek2020, Markowitz2016}, the decorrelation speed of the OCT signal \cite{AbdElSadek2020,ElSadek2021}, and the time-frequency spectrum of the OCT signal \cite{Muenter2020}.

The DOCT signal is a measure of the time property of the OCT signal, but it does not reflect the intracellular/intratissue motion directly.
This can be understood by considering the following example.
During the OCT signal formation process, moving scatterers within the tissue modulate the scattered light field. 
The light component that is scattered from each scatterer corresponds to a phasor contribution to the complex OCT signal, and a complex OCT signal at a single point in an OCT image is the summation of the multiple phasors.
The OCT signal intensity  is computed as an absolute-squared value of the complex OCT signal, i.e., as the product of the complex OCT and its complex conjugate.

In this process, the phasor contributions mutually interact and form a speckle pattern \cite{zhou_unified_2021, Tomita2023}.
Assume that a scatterer in the sample moves with uniform speed.
Although the scatterer’s motion is a linear motion, the phase of the corresponding phasor oscillates with an oscillation frequency that is proportional to the axial component of the motion velocity.
More complex cases can be exemplified by considering two scatterers that are moving at the same uniform speed but in different directions, i.e., at different axial speeds.
First, the phasors that correspond to these scatterers have different phase-modulation frequencies, even though they are moving with the same speed (i.e., absolute velocity).
Second, these phasors interact during the intensity-signal formation process and it results in additional frequency components in the intensity signal fluctuation that do not relate directly to either scatterer’s velocity.

To date, the relationship between DOCT signals and the individual scatterer motions in the tissue has not been fully identified.
As a result, the dynamic OCT contrast cannot really be interpretable.

In this paper, we present a dynamic OCT simulation framework (i.e., a dynamic OCT simulator) that enables numerical investigation of the intracellular/intratissue-motion-to-DOCT-signal relationship.
This simulation framework consists of mathematical models of the intracellular/intratissue scatterer motions that can be used to generate a numerical distribution of the scatterers, and a numerical OCT simulator that generates the OCT intensity pattern (i.e., the speckle pattern) from the scatterer distribution.
By generating a time sequence of the numerical distributions of the dynamic scatterers using the scatterer motion models, the numerical OCT simulator can then generate the time sequence of the dynamic speckle patterns.
In addition, the OCT simulator is equipped with a physically accurate mathematical model of the OCT noise that reflects the spatial properties of shot noise, relative intensity noise (RIN), and non-optical noise \cite{Seesan2024}.
Therefore, the effects of noise in DOCT imaging can also be investigated using the simulation framework.
The performance of our simulation framework is demonstrated by conducting two example studies to investigate the relationships between the DOCT signals and the scatterers' motion speed, noise, and dynamic-scatterer ratios (i.e., the occupancy of the dynamic scatterers).
An open source implementation of the simulator written in Python 3.10.13 with NumPy v1.24.3 and CuPy v12.3.0, and can be found at \url{https://github.com/ComputationalOpticsGroup/COG-dynamic-OCT-contrast-simulation-library} \cite{GitHubDoctSim}.

\section{Simulation framework}
\subsection{Modeling of intracellular/intratissue dynamics}
\label{sec:scattererModel}
According to a literature survey, seven intracellular/intratissue activity types have been considered including both active \cite{Arcizet2008,Papin2005,Koumakis2014} and passive \cite{Arcizet2008,Papin2005,Solomon1988} transports (for proteins), cyclosis (for organelles) \cite{Goldstein2015}, the motion of dissociated floaters (for organelles) \cite{Li2018}, intracellular jiggling (for organelles) \cite{Kendal1983}, cell migration (for cells) \cite{Wortel2021}, and blood flow (for cells) \cite{Hoque2018,Arend1991}.
For each of these activities, one of three types of intracellular and intratissue scatterers, including proteins, organelles and cells, are considered.

In addition to these activity types, the typical diameters and masses \cite{Ripple2012,Fang2003,Koopman2005,Phillips2012,Persons1929,Emerman1979}, the speed and duration of motion, and/or the diffusivity of each of the scatterer types are summarized in Table\@ \ref{Table:parameterSurvey}.

\begin{figure}
	\centering\includegraphics{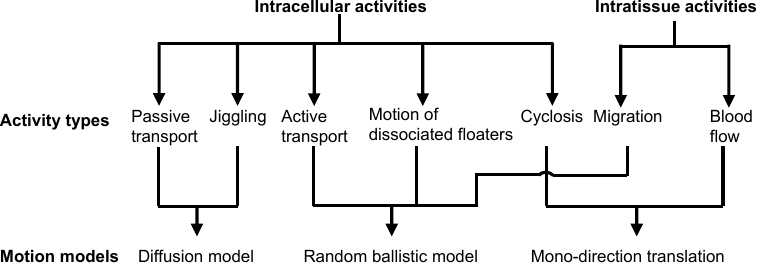}
	\caption{%
		Summary of intracellular/intratissue activities and mathematical models of the intracellular/intratissue motions.
		Seven activity types are included, and each activity type is expressed using one of three types of motion models.
	}
  	\label{fig:classification}
\end{figure}
In our simulation framework, the scatterer motions related to the intracellular and intratissue activities described above are modeled mathematically using three motion types, including random-ballistic motion, diffusion, and mono-directional translation, as summarized in Fig.\@ \ref{fig:classification}.

\begin{figure}
\centering\includegraphics{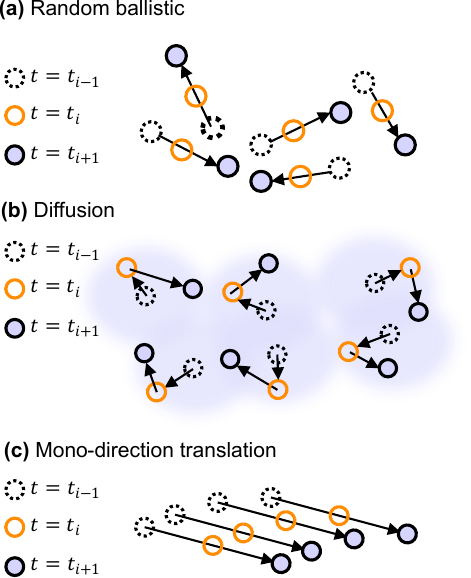}
\caption{
		Schematic illustrations of the three scatterer motion models.
		(a) In the random ballistic model, all scatterers are move linearly with the same speed, but the directions of motion are randomly defined for each scatterer.
		(b) In the diffusion model, the scatterers follow the random walk motion (i.e., Brownian motion).
		(c) In the mono-directional translation model, all scatterers travel linearly in the same direction at the same speed.
		$i$ is the time point index, and $t_{i-1}$, $t_{i}$, and $t_{i+1}$ represent three time points.
	}
	\label{fig:Models}
\end{figure}

\begin{table}[]
	\caption{%
		Summary of intracellular/intratissue dynamic activities and their properties.
	}
	\centering\includegraphics{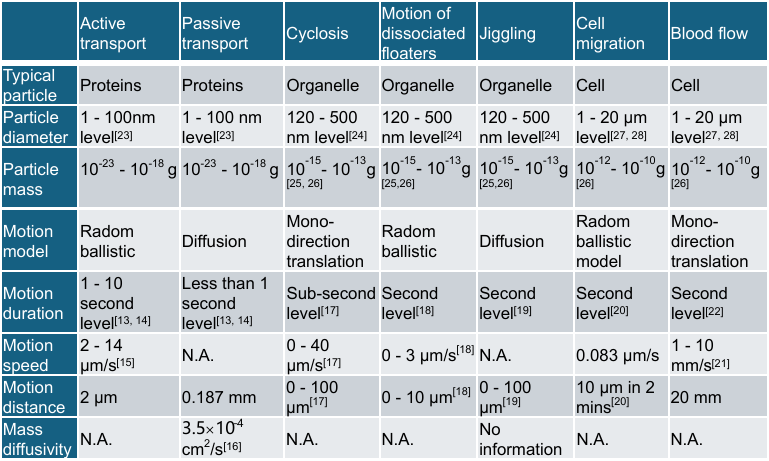}
	\label{Table:parameterSurvey}
\end{table}
The random-ballistic motion models active transport, the motion of dissociated floaters, and cell migration. 
In this model, all scatterers are in uniform linear motion at the same speed (i.e., the absolute velocity) but move in different randomly defined directions, as illustrated in Fig.\@ \ref{fig:Models}(a). 

This motion is described mathematically as  
\begin{equation}
	\boldsymbol{r}_{j, i+1} = \boldsymbol{r}_{j, i} + v\boldsymbol{e}_j(t_{i+1}-t_i),
	\label{eq:randomModel}
\end{equation}
where $\boldsymbol{r}_{j, i}$ represents the three-dimensional (3D) location of the $j$-th scatterer at the $i$-th time point ($t_i$), $v$ is the absolute velocity of the motion, $\boldsymbol{e}_j$ represents the unit velocity vector of the motion of the $j$-th scatterer, and $(t_{i+1}-t_i)$ is the time interval between the $i$-th and $(i+1)$-th time points.

The diffusion models passive transport and jiggling.
In this model, the scatterers are assumed to follow the random walk motion, in which the displacement of each scatterer is random and conforms to a zero-mean normal distribution, as illustrated in Fig.\@ \ref{fig:Models}(b).
The variance of the normal distribution is related to the diffusion coefficient as follows: 
\begin{equation}
\langle(\boldsymbol{r}_{j, i+1}-\boldsymbol{r}_{j, i})^2\rangle_j = 6D(t_{i+1}-t_i),
\label{eq:diffusionModel}
\end{equation} 
where the left hand side of the equation is the variance with $\langle \rangle_j$ is the ensemble average over all scatterers and $D$ is the diffusion coefficient.

For the numerical simulations, the position of each scatterer can be updated using
\begin{equation}
	\boldsymbol{r}_{j, i+1} = \boldsymbol{r}_{j, i}+ \mathcal{N}\left(\mu = \mathbf{0}, \sigma_{x,y,z}^2 = 2D(t_{i+1}-t_i)\right),
	\label{eq:diffusionUpdate}
\end{equation} 
where $\mathcal{N}$ is a zero-mean 3D normal distribution.
The variances for all directions ($\sigma_{x,y,z}^2$) are equally $2D(t_{i+1}-t_i)$, and hence, the variances of the absolute displacement $\mathbf{\sigma}^2$ becomes $6D(t_{i+1}-t_i)$ (that is $\sigma_x^2 + \sigma_y^2 + \sigma_z^2$).

The mono-directional translation models cyclosis and blood flow.
In this model, all scatterers follow a uniform linear motion with the same velocity amplitude and direction, as illustrated in Fig.\@ \ref{fig:Models}(c) and modeled mathematically as 
\begin{equation}
	\boldsymbol{r}_{j, i+1} = \boldsymbol{r}_{j, i} + \boldsymbol{v}(t_{i+1}-t_i),
	\label{eq:flowModel}
\end{equation} 
where $\boldsymbol{v}$ denotes the velocity vector of the motion.

\subsection{OCT signal formulation models}
\label{sec:cleanOctFormation}
We propose two types of OCT signal formation model.
The first is the single-point-signal formation model, which models the temporal pattern of the OCT signals at a single point in the image.
Although this model cannot simulate the spatial pattern of the OCT signal, it is computationally time-efficient.
The second is the 3D-OCT-volume formation model, which can be used to generate the time sequences of volumetric OCT speckle patterns.
In addition, this model can provide a physically accurate emulation of the spatial pattern of the noise.
Both models are based on a simplified version of Tomita's numerical sample model, known as the dispersed scatterer model (DSM) \cite{Tomita2023}.
In the original DSM, the sample was modeled using randomly distributed, infinitely small scatterers that were embedded in a medium with a spatially slowly-varying refractive index.
In our simulator, this model was simplified by assuming that the medium has a spatially uniform refractive index.

\subsubsection{Single-point-signal formation model}
\label{sec:spsModel}
In the single-point-signal formation model, the complex OCT signal at a specific point in the image $S(t_i)$ is expressed as the sum of the electric-field contributions from the individual scatterers as
\begin{equation}
	\label{eq:single complex OCT}
	S(t_i) \propto \sum_{j=0}^{N_s-1} {b_j 
		\exp\left[i2\pi\frac{2z_{j,i}}{\lambda} + i\phi \right]} 
	\textcolor{purple}{\exp\left[-\frac{1}{2}\left(\frac{x_{j,i}}{\sigma_x}\right)^2-\frac{1}{2}\left(\frac{y_{j,i}}{\sigma_y}\right)^2-\frac{1}{2}\left(\frac{z_{j,i}}{\sigma_z}\right)^2\right]},  
\end{equation}
where $j$ and $i$ are the indexes of the scatterer and the time, respectively, and $N_s$ represents the total number of scatterers in the sample.
Here, each term in the summation represents the phasor contribution from each scatterer to a single point in the image.
$b_j$ is a real coefficient that accounts for the amplitude of the reflectivity of the scatterer, which is mainly determined based on the scattering potential [see the discussion around Eq. (1) in Ref.\@ \cite{Tomita2023}].
In our corresponding demonstration during the later sections of this work, we assume that $b_j = 1$ for simplicity.
The first exponential component in the equation is the phase of the phasor, which is defined by the depth position of the scatterer $z_j$, the center wavelength of OCT $\lambda$, and a constant phase offset $\phi$ that is defined by the depth offset between the reference and probe arms.
Because $\phi$ remains constant for all scatterers, it does not influence the OCT signal intensity.

The second exponential term (which is highlighted in red) represents the weight of the contribution of each scatterer as defined by the 3D Gaussian point spread function (PSF).
$\sigma_x, \sigma_y$, and $\sigma_z$ denote the widths of the amplitude PSF, as defined by the standard deviation of the Gaussian function.
The $1/e^2$-width and the full width at half maximum (FWHM) of the PSF are related to the corresponding standard-deviation width ($\sigma$) as 
\begin{equation} 
	\label{eq:sigma-e2}
	1/e^2 \mathrm{-width} = 2\sqrt{2} \sigma,
\end{equation}
\begin{equation} 
	\label{eq:sigma-fwhm}
		\mathrm{FWHM} = {2\sqrt{\ln 2}} \sigma.
\end{equation}

Finally, the OCT signal  intensity at time point $t_i$ is defined as $I(t_i) = S(t_i)S^*(t_i)$.

\subsubsection{3D-OCT-volume formation model}
\label{sec:3dOctModel}
The 3D-OCT-volume formation model is similar to that presented in Ref.\@ \cite{Seesan2021,Seesan2024}, except that our model considers the motion of the scatterers.
In this model, we first generated a 3D scatterer field.
Here, each scatterer is expressed using an infinitely small complex point as 
\begin{equation}
	P_{j}(x,y,z; t_i)
	= b_{j}
	\exp\left[i2\pi\frac{2z_{j,i}}{\lambda} + i\phi \right] 
	 \delta(x-x_{i,j},y-y_{i,j},z-z_{i,j}),
	\label{eq:3Dreflectivity}
\end{equation}
where $j$ and $i$ are the indexes of the scatterer and the time, respectively.
$b_{j}$ and the exponential part are the same as the corresponding components of the single-point-signal formation model, and the delta function represents the scatterer position.
In the simulations presented in the later sections, the phase offset $\phi$ is ignored without losing generality, and $b_j$ was set at 1 for simplicity.

The 3D scatterer field is then defined as the collection of all scatterers as
\begin{equation}
	P_{all}(x,y,z, t_i) = \sum_{j=0}^{N_s-1} P_{j}(x,y,z; t_i).
	\label{eq:3DscatterField}
\end{equation}

The PSF is assumed to be a 3D Gaussian function where
\begin{equation}
	\mathrm{PSF}(x,y,z) \propto \exp{- \left[
		4\left(\frac{x}{\Delta x}\right)^2
		+ 4\left(\frac{y}{\Delta y}\right)^2
		+ 2\ln{2}\left(\frac{z}{\Delta z}\right)^2
		\right]},
    \label{eq:PSF}
\end{equation} 
where $\Delta x$ and $\Delta y$ are the lateral resolutions, which are defined as the $1/e^2$-width of the squared amplitude of the PSF, and $\Delta z$ is the axial resolution defined by the FWHM of the squared amplitude.

The 3D complex OCT signal is then computed by convolving the 3D PSF with the scatterer field as 
\begin{equation}
	S(x,y,z; t_i) = P_{all}(x,y,z, t_i) * \mathrm{PSF}(x,y,z),
    \label{eq:3DComplexOCTSignal}
\end{equation} 
where * indicates the convolution.
The 3D OCT intensity distribution, i.e., the 3D speckle, is finally obtained as $I(x,y, z; t_i) = S(x, y, z; t_i)S^*(x, y, z; t_i)$.

In the numerical implementation, both the scatterer field and the PSF are voxelized, with each scatterer occupying a single voxel. 
The convolution of Eq.\@ (\ref{eq:3DComplexOCTSignal}) is computed using a fast Fourier transform (FFT)-based method.

\subsection{Noise model}
\label{sec:noiseModel}
OCT imaging suffers from three noise types, including shot noise, relative intensity noise (RIN), and non-optical noise (no-noise).
The amplitudes of these three noise types have different light intensity dependences\cite{Leitgeb2003}.
Specifically, the amplitudes of shot noise and RIN are proportional to the amplitude (i.e., the square-root of the intensity) and the intensity of the light, respectively, while the no-noise amplitude is independent of the light intensity.
Although these noise types are all Gaussian noises, their different light intensity dependences cause them to have different spatial correlation properties, as follows \cite{Seesan2024}.

In the wavenumber domain, the total noise ($N_{all}$) is the sum of these three noise types and is given by
\begin{equation}
	N_{all}(k) = c_1 \sqrt{S_p(k)}N_{\mathrm{sh}}(k) + c_2 S_p(k) N_{\mathrm{RIN}}(k) + c_3 N_{\mathrm{no}}(k),
    \label{eq:NoiseModel_kDomain}
\end{equation}
where $k$ is the wavenumber and $S_p(k)$ represents the power spectrum of the light source.
$N_{\mathrm{sh}}$, $N_{\mathrm{RIN}}$, and $N_{\mathrm{no}}$ are all normally distributed random variables that correspond to the shot noise, the RIN, and the non-optical noise, respectively.
All these random variables are assumed to have zero means and the same standard deviations, but they are independent of each other.
In addition, these random variables are fully decorrelated along $k$.
The constants $c1$, $c2$, and $c3$ are proportionality constants that control the magnitude of each noise type.

The depth domain expression for the noise is given by the inverse Fourier transform of Eq.\@ (\ref{eq:NoiseModel_kDomain}) as 
\begin{equation}
    \widetilde{N_{all}}(z) = c_1 \ift{\sqrt{S_p(k)}}(z) *\widetilde{ N_{\mathrm{sh}}}(z) + c_2 \ift{S_p(k)}(z) * \widetilde{N_{\mathrm{RIN}}}(z) + c_3 \widetilde{N_{\mathrm{no}}}(z),
     \label{eq:NoiseModel_zDomain}
\end{equation}
where $z$ is the depth and is a Fourier pair of $k$.
Both $\ift{\quad}$ and tilde ($\widetilde{\quad}$) denote the inverse Fourier transform operation.
The Fourier transforms of the random variables, i.e., $\widetilde{N_{\mathrm{sh}}}(z)$, $\widetilde{N_{\mathrm{RIN}}}(z)$, and $\widetilde{N_{\mathrm{no}}}(z)$, are zero-mean complex Gaussian noises and they are independent of each other.

Note that although each of these random variables are fully decorrelated along the depth ($z$) direction, the convolution with the $S_p(k)$-related parts causes spatial correlation of the noises.
In general, $\sqrt{S_p(k)}$ is wider than $S_p(k)$ and thus $\ift{\sqrt{S_p(k)}}(z)$ is narrower than $\ift{S_p(k)}(z)$.
When this point is considered, Eq.\@ (\ref{eq:NoiseModel_zDomain}) indicates that the RIN has the largest correlation distance along the $z$ direction, whereas the no-noise is fully decorrelated.

Our numerical simulation framework can generate numerical noises that follow these spatial properties.
Specifically, we numerically generated the each type of noise in $k$-domain by Eq.\@ (\ref{eq:NoiseModel_kDomain}), and applied an inverse FFT to obtain the noise in the $z$-domain, which corresponds to Eq.\@ (\ref{eq:NoiseModel_zDomain}). 
Subsequently, the noise is rescaled independently for each type, and hence, we can control the energy of each noise independently  and also can control the energy balance.
The complex OCT signal measured under the noise (noisy complex OCT) can then be obtained by adding the numerical noise to the noise-free complex OCT signal to give 
\begin{equation}
    S'(x,y,z; t_i) = S(x,y,z; t_i) + \widetilde{N_{all}}(z; t_i),
     \label{eq:measuredSignal}
\end{equation}
where $S'(x,y,z; t_i)$ is the noisy complex OCT and $S(x,y,z; t_i)$ is the noise-free complex OCT obtained by the method described in Section \ref{sec:cleanOctFormation}.
Here, the noise is written as a function of time (where $t_i$ is the time of the $i$-th simulation time point).
For each time point, the numerical noise can be generated independently because the noise is fully decorrelated with respect to the time, 

Finally, the OCT intensity, i.e., the speckle pattern under the noise, is computed as the square of the intensity of the noisy complex OCT as $I(x,y, z; t_i) = S'(x, y, z; t_i)S'^*(x, y, z; t_i)$.

\subsection{Simulation flow}
\subsubsection{Simulation flow of the single-point-signal formation}
\label{sec:pointSimulationFlow}
The numerical generation of OCT signal based on the single-point-signal formation model (Section \ref{sec:spsModel}) is straightforward.
Specifically, it is not necessary to discretize the scatterer positions and to consider discrete 3D fields.
Instead, the positions of the scatterers and their time increments are computed directly using Eqs.\@ (\ref{eq:randomModel}), (\ref{eq:diffusionUpdate}), and/or (\ref{eq:flowModel}).
In addition, the complex OCT signal is computed using Eq.\@ (\ref{eq:single complex OCT}) without discretizing the space.

It should be noted that the initial distribution of the scatterers (i.e., the seeding-space size) should be sufficiently wider than the PSF.
If the seeding-space size is too small, then after a certain time has elapsed, some scatterers that were initially outside the seeding space may enter the PSF region.
But their contributions will be ignored in the simulation because they have not been numerically embodied.
Here, we refer to these scatterers as ignored scatterers.
For example, in a recent publication by the authors, a simulation based on the random-ballistic model was used to analyze the properties of newly introduced DOCT algorithms \cite{Morishita2024a}.
In that work, the seeding-space size was set to be $5\sigma + 2 vT$ for each side, where $\sigma$ represents the standard-deviation size of the amplitude PSF, $v$ is the absolute velocity (speed) of the scatterer, and $T$ is the total simulated time duration. 
Under these conditions, the ignored scatterers cannot reach the area covering 99.88\% of the light energy of the PSF.
Further details of the above can be found elsewhere (e.g., in Section 3.1.3 and in Supplementary section S1 of Ref.\@ \cite{Morishita2024a}).

\subsubsection{Simulation flow of 3D-OCT-volume formation}
\label{sec:3DsimualtionFlow}
\begin{figure}
	\centering\includegraphics[width=13cm]{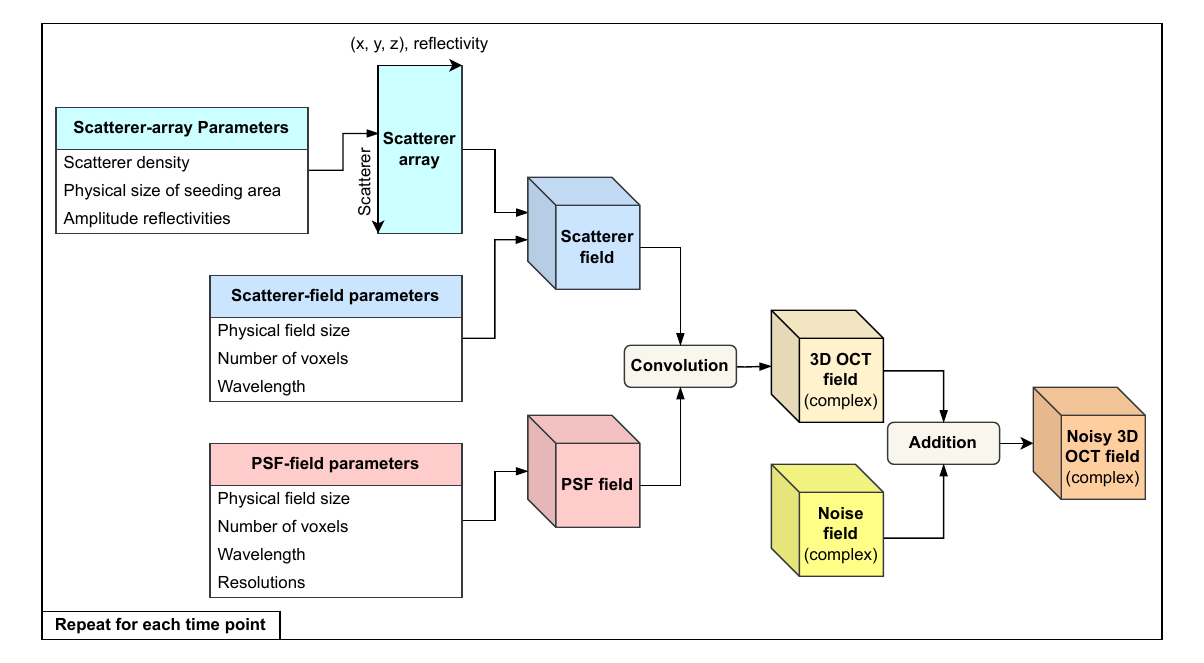}
	\caption{%
		The flow to generate 3D OCT volume, i.e., 3D speckle pattern, using the 3D-OCT-volume formation model.
		The details of the flow is described in Section \ref{sec:3DsimualtionFlow}.
	\label{fig:3dModelFlow}
	}
\end{figure}
The simulation flow using the 3D-OCT-volume formation model is schematically summarized in Fig.\@ \ref{fig:3dModelFlow}.
In this process, we first generate two 3D complex numerical fields: the scatterer field and the PSF field. 
The scatterer field is a 3D complex array that holds the complex reflectivity distribution of the scatterers.
To generate the scatterer field, we first generate (or seed) the scatterers, where each scatterer is represented by a set of four parameters, including its 3D position ($x$, $y$, $z$) and real amplitude reflectivity.
Additionally, the set of scatterers (called the scatterer array) is represented by a 2D numerical array, in which each row corresponds to each scatterer and each column represents each of $x$, $y$, $z$, and the amplitude reflectivity.
To generate the scatterer array, we must define several parameters, including the scatterer density, the physical size of the seeding area, and the amplitude reflectivities.
In our specific initialization, the scatterers are spatially randomly seeded.

The scatterer field is then generated from the scatterer array and other parameters, including the physical field size, the number of voxels in the field, and the wavelength of the OCT system.
Here, the scatterer field size is set to be identical to the seeding area of the scatterer array.
Each voxel in the scatterer field represents the complex reflectivity of the scatterer(s), where the phase of the reflectivity is affected by the scatterer’s depth position ($z$) and the wavelength through Eq.\@ (\ref{eq:3Dreflectivity}).
If there are multiple scatterers within a voxel, the voxel value is defined as the sum of the phasors of all scatterers.
Note that the scatterer field corresponds to $P_{all}$ from Eq.\@ (\ref{eq:3DscatterField}).

The other numerical field is the PSF field, which is a complex 3D array that holds the complex distribution of the PSF.
The PSF field is generated using several parameters, including the physical field size, the voxel numbers, the probe wavelength, and the resolutions.
The resolutions in this case are 3D resolutions, i.e., they are three real parameters that define the resolutions along the $x$, $y$, and $z$ directions.
The field size and the voxel numbers are the 3D physical field size and the number of voxels for each of the $x$, $y$, and $z$ directions.
In our open source implementation, the field sizes and the voxel numbers for the PSF field and those for the scatterer fields are identical.
Although this initially causes a large void space in the PSF field, it later eases the convolution between the two field types.
(See the next paragraph for further details.)
Here, we assume that the PSF is a 3D Gaussian [Eq.\@ (\ref{eq:PSF})].
Although we assumed that the PSF is both defocus-free and aberration-free in our current implementation, these factors can be considered easily by introducing distorted amplitude and phase components to the PSF.

After the two numerical fields (i.e., the scatterer field and the PSF field) have been generated, we generate the OCT field, which is a 3D complex numerical array that represents the volumetric complex OCT signal, by convolving the two numerical fields three-dimensionally.
In our specific implementation, the convolution was computed by fast-Fourier-transform (FFT) based algorithm. 
And hence, physical field sizes and the voxel numbers of the scatterer- and the PSF-fields must be identical.
Note here that, due to the FFT-based convolution operation, when a scatterer is at the very border of the scatterer field convolved with the PSF, some part of the signal will be aliased into the opposite border of the field.
The periphery of the OCT field up to at least the approximate size of the PSF is affected by the aliasing and thus should be excluded from further use (in the analyses).
In addition, as we discuss in the following paragraphs, the scatterer position is updated at each time step increment.
This causes some virtual scatterers that lie outside the seeding area (which are called void scatterers) to enter the scatterer field.
Because these scatterers are not numerically embodied (i.e., seeded), the contributions of the void scatterers are erroneously ignored.
As a result, a peripheral portion of the OCT field that is even larger than the PSF size in some cases becomes unreliable and should thus be discarded from further analyses.
We must therefore set the scatterer field size (and thus set the sizes of the PSF and OCT fields) to be large enough to maintain a sufficient field volume size for the subsequent steps in the analysis. 

The steps described above gave us the 3D OCT field at the initial time point.
In the next step, the scatterer position array must be updated for the next time point using one of the scatterer motion models [Section \ref{sec:scattererModel}, Eqs.\@ (\ref{eq:randomModel}), (\ref{eq:diffusionUpdate}), and (\ref{eq:flowModel})].
For this scatterer position update, we use a set parameter for either the scatterer speed (for both random-ballistic motion and mono-directional translation) or the diffusion coefficient  (for diffusion).
Then, a new OCT field is computed using the same process flow with the first time points.
Finally, the full OCT time sequence is computed by repeating this process until the last time points are reached.
For these computation steps, we used two setting parameters, comprising the time interval $\Delta t$ and the number of volumes in the sequence.

Notably, we can consider combinations of scatterers with different motion types by one of two means.
In one method, a portion of the scatterers within a scatterer array (a set of scatterers) are updated using one of the motion models, while another portion is updated by applying another motion model.
The scatterer field is then computed from the updated scatterer array and is convolved with the PSF field to generate the OCT field required.
In addition, if we do not update some portion of the scatterer array, i.e., if we use the ``no-motion model,'' we can then simulate the case with an arbitrary dynamic scatterer ratio.
In the other method, an OCT field is generated using a partial scatterer array that is updated using a particular motion model (including the no-motion model), and another OCT field is generated using another partial scatterer array.
The final OCT field is then obtained via complex summation of these partial OCT fields.
In the examples presented in Section \ref{sec:dsrStudy}, we used the latter method.

To make the simulation conditions realistic, noise can be added to the complex OCT field.
For this process, a noise field, which is a 3D complex array that has the same size and the same voxel numbers as the OCT field, is computed using Eqs.\@ (\ref{eq:NoiseModel_kDomain}) and \@(\ref{eq:NoiseModel_zDomain}).
Here, constant magnitudes are set for each noise type ($c1$, $c2$ and $c3$)  to achieve a particular signal-to-noise ratio (SNR) for each noise type.
The noise field is then added to the OCT field.
Further details of the noise incorporation procedure can be found elsewhere \cite{Seesan2024}.


\subsection{DOCT algorithms}
\label{sec:DOCTs}
The simulator is equipped with DOCT methods that compute the DOCT values from the simulated time sequences of the OCT signals.
These DOCT methods include the logarithmic intensity variance (LIV), the OCT correlation decay speed (OCDS), and the amplitude-spectrum-based method.

LIV is a DOCT contrast that is sensitive to the occupancy of the dynamic scatterers \cite{Morishita2024a} and is defined as the time variance of the logarithmically (dB) scaled OCT intensity \cite{AbdElSadek2020,ElSadek2021} as
\begin{equation}
\begin{split}
\mathrm{LIV}(x,z) = &
\frac{1}{N} \sum_{i = 0}^{N-1}
\left(I_{\mathrm{dB}}(x,z,t_i) - \overline{I_{\mathrm{dB}}} \right)^2,
\end{split}
\label{eq:LIV}
\end{equation}
where $I_{\mathrm{dB}}(x, z, t_i)$ is the $i$-th frame of a dB-scaled OCT frame sequence, where $i = 0, 1, 2, \cdots, N-1$, and $N$ is the number of frames.
$\overline{I_{\mathrm{dB}}}$ represents the average of all the frames.

OCDS is a DOCT contrast that is sensitive to the speed of the dynamics, and OCDS is defined as the slope of the autocorrelation function of the time sequence of the OCT signals within a specific delay time range\cite{AbdElSadek2020,ElSadek2021}. 
Here, the autocorrelation function is computed as 
\begin{equation}
\rho_A(\tau; x, y) =
\frac{\mathrm{Cov}\left[I_{dB}(x,z,t_i),\, I_{dB}(x,z,t_i+\tau)\right]}
	{\mathrm{Var}\left[I_{dB}(x,z,t_i)\right] \mathrm{Var}\left[I_{dB}(x,z,t_i+\tau)\right]},
\label{eq:corr_coeff}
\end{equation}
where $\tau$ is the delay time, and $\mathrm{Cov}$ and $\mathrm{Var}$ represent the covariance and the variance, respectively.

The amplitude spectrum is the amplitude of the time spectrum of the OCT signal sequence, where the spectrum is computed by discrete Fourier transformation of the amplitude of the OCT signal sequence.
For the graph plots shown in Section \ref{sec:exampleStudy}, the amplitude spectra are averaged over the 3D speckle volume.
Along with the graph plot, an amplitude-spectrum-based DOCT (AS-DOCT) image is also computed using a method similar to that demonstrated by M\"unter \etal \cite{Muenter2020, Muenter2021}.
In this method, each amplitude spectrum is divided into three frequency sub-bands and is then integrated over each sub-band.
This operation gives three OCT images corresponding to the three spectral bands.
Each of the OCT images are logarithmized, scaled into a [0, 1] range, and assigned to red, green, and blue channels, and a pseudo-color AS-DOCT image is then obtained.

\section{Example studies}
\label{sec:exampleStudy}
To showcase our simulation framework, we present two example studies here.
The first study is the ``fully dynamic study,'' in which all the scatterers are fully dynamic. 
In this study, we investigate the motion parameters (scatterer speeds and diffusion coefficients) and the SNR dependences of the DOCT signals.
The second is called the ``dynamic-scatterer-ratio (DSR) study,'' in which the DOCT values of partially dynamic samples are investigated.
In the partially dynamic samples, some of the scatterers follow one of the three motion types, whereas the other scatterers are static.

\subsection{Fully dynamic study}
\label{sec:fullyDyanmicStd}
In this simulation scenario, the sample is assumed to be fully dynamic.
In other words, all the scatterers move by following one of the three motion models.
Here, we perform simulations with different motion parameters, i.e., different scatterer speeds or diffusion coefficients, and analyze their influence on the DOCT values.

\subsubsection{Study protocol of the fully dynamic study}
\label{subsubsec: protocol}
Two time protocols were used in this study.
For the LIV and OCDS analyses, each time sequence consists of 32 frames with an inter-frame time interval of 204.8 ms, and the total time to be simulated is 6.35 s.
For the analysis of the amplitude spectrum, each time sequence consists of 150 frames with an inter-frame time interval of 9.33 ms, and the total simulated time is 1.39 s.
The former approach is similar to the experimental protocol used to perform 3D DOCT measurements previously by the authors \cite{ElSadek2021}, while the time interval and the frame number of the latter method are identical to those used by M\"unter \etal \cite{Muenter2021}.

The simulation was performed with the 3D-OCT-volume formation model (Section \ref{sec:3dOctModel}), with scatterer field dimensions of 345 \um ($x$) $\times$ 345 \um ($y$) $\times$ 874 \um ($z$) and OCT field dimensions of 195 \um ($x$) $\times$ 195 ($y$) \um$\times$ 724 \um ($z$).
The OCT field consists of 100 $\times$ 100 $\times$ 100 voxels and thus the volxel separations are 1.95 \um (lateral) and 7.24 \um (axial).

In the OCT system specifications, the probe wavelength is 1,310 nm.
The axial and lateral resolutions were set to be 14 \um and 10.6 \um in their FWHM values, respectively, which are 23.7 \um and 18 \um in $1/e^2$ widths, respectively.
These parameters are the system parameters of the swept-source OCT used in our previous DOCT studies \cite{AbdElSadek2020,ElSadek2021}.

The parameters for the scatterers and their motions were defined as follows. 
The reflectivity was set as 1 for all scatterers.
For both the random-ballistic and mono-directional translation models, all scatterers are moved with the same speed, which varied from 4.5 to 5400 nm/s over several simulation trials.
For the mono-directional translation model, the flow angle was set to be at 45 degrees to the depth axis ($z$-axis).
For the diffusion model, all scatterers shared the same diffusion coefficient during each simulation trial, but its value varied from 20 $\times 10^{-18}$ m$^2$/s to 30 $\times 10^{-12}$ m$^2$/s.
The scatterer density was set at 0.055 scatterers/\cbum, which is a typical scatterer density for tumor spheroids that was identified using a neural-network-based scatterer density estimator \cite{Seesan2021}.  

We added noise generated with the non-optical noise model (Section \ref{sec:noiseModel}) to realize several SNRs, which ranged from 0 dB to 40 dB.

The LIV, the OCDS, and the amplitude spectrum were then computed from the simulated OCT sequence.
For each set of simulation parameters, eight simulation trials were performed for both the LIV and the OCDS.
For the amplitude spectrum analysis, only six trials were used because each trial required 150 frames and hence required very intense computation, whereas the LIV and the OCDS only used 32 frames.
The delay range for the OCDS was set at [204.8 ms, 1,228.8 ms].

\subsubsection{Dynamic speckle examples}
\begin{figure}
	\centering\includegraphics{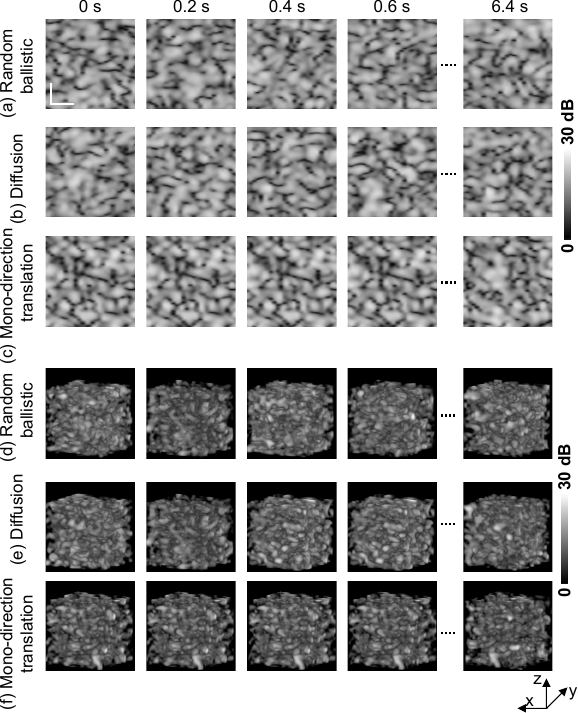}
	\caption{%
		Example time sequences of simulated speckles.
		The scatterers here follow one of the random ballistic, diffusion, or mono-directional translation models, as indicated on the left of each of the images.
		The first three rows show 2D cross-sections extracted from the 3D volumes and the last three rows show volume rendering of the 3D speckle.
		The speckle motion can also be seen in the corresponding videos: Visualization 1 (random ballistic), Visualization 2 (diffusion), and Visualization 3 (mono-directional translation).
	}
	\label{fig:speckleTimeSequence}
\end{figure}
Examples of the time sequences of the simulated speckle patterns of the three different scatterer motion models are shown in Fig.\@ \ref{fig:speckleTimeSequence}.
The scatterer velocity was set at 4.5 \um /s for both the random-ballistic and mono-directional translation motions.
The angle of the mono-directional translation was set at 45 degrees to the optical axis.
The diffusion coefficient for the diffusion model was 21 $\mu$m$^2$/s. 
We did not add noise in these specific cases.

The upper three rows show representative cross-sections that were extracted from the simulated 3D volumes, where the scale bar represents 50 \um.
The lower three rows show the volume rendered 3D speckle patterns.
The corresponding media (i.e., Visualizations 1 to 3) show time-lapse movies of the 3D speckle volumes.
The volume rendered animation was created using open-source visualization software (Open Chrono-Morph Viewer) \cite{Faubert2024}.

As the movies clearly show, in the mono-directional translation case, the speckle pattern is fully preserved but only travels in a single direction (Visualization 3).
In contrast, both the random ballistic motion (Visualization 1) and diffusion (Visualization 2) cases caused rapid and random changes in the speckle pattern.

\subsubsection{Results of the fully dynamic study}
In the systematic investigation, we performed multiple trials (eight each for LIV and OCDS, and six for the amplitude spectrum) for each simulation configuration.
The graphs plotted in this section show the mean values for each of the trials, and the error bars indicate the standard deviations, although they are negligibly small throughout.

\begin{figure}
	\centering\includegraphics{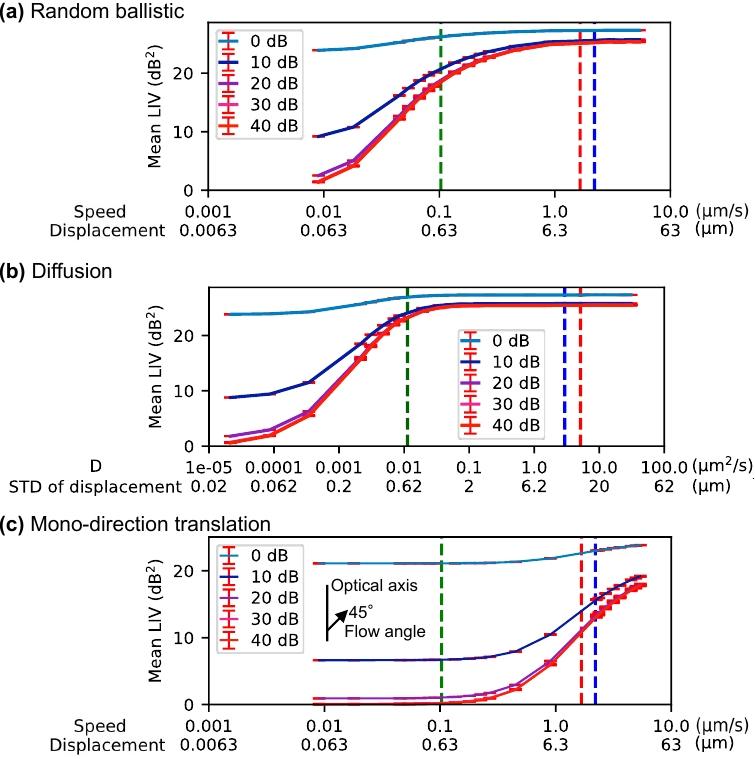}
	\caption{%
		Mean LIV values shown as motion parameters (the scatterer speed and the diffusion coefficient). 
		The error bars indicate the standard deviations among the eight trials.
		The first label on each horizontal axis represents the motion parameters, and the second label represents the corresponding travel distance during the full simulation time.
		The vertical dashed lines represent half the wavelength (green) and the lateral (red) and axial (blue) resolutions.
		Plots (a) to (c) represent (a) the random ballistic motion, (b) diffusion, and (c) mono-directional translation.
		The mean LIV values were computed for five different SNR configurations.
		The LIV was found to increase as the relevant motion parameter value increased.
	}
	\label{fig:LIVNoise}
\end{figure}
Figure \ref{fig:LIVNoise} summarizes the results for the LIV.
In this case, the first label on the horizontal axis represents the speed parameters (i.e., the scatterer speed or the diffusion coefficient) and the second label represents the maximum displacement of the scatterers during the simulation time (for the random ballistic motion and the mono-directional translation) or the standard deviation of the maximum displacement (for the diffusion).
The dashed horizontal lines correspond to the second label and indicate half the wavelength (i.e., the full wavelength for double-pass; green), and the FWHM resolutions of the lateral (red) and axial (blue) directions.
The same descriptions apply for all plots shown in the later figures in this paper.
For all the motion models, the mean LIV increases as the scatterer speed or the diffusion coefficient (i.e., the motion parameters) increases, but the increasing rate becomes lower for high motion parameter values.
Therefore, the LIV becomes less sensitive to the motion parameters as the motion parameter values increase.
The LIV value becomes less sensitive to the motion parameters when the SNR decreases.
The LIV takes higher values for lower SNR values, and becomes almost constant, regardless of the motion parameters if the SNR is lower than 10 dB.

\begin{figure}
	\centering\includegraphics{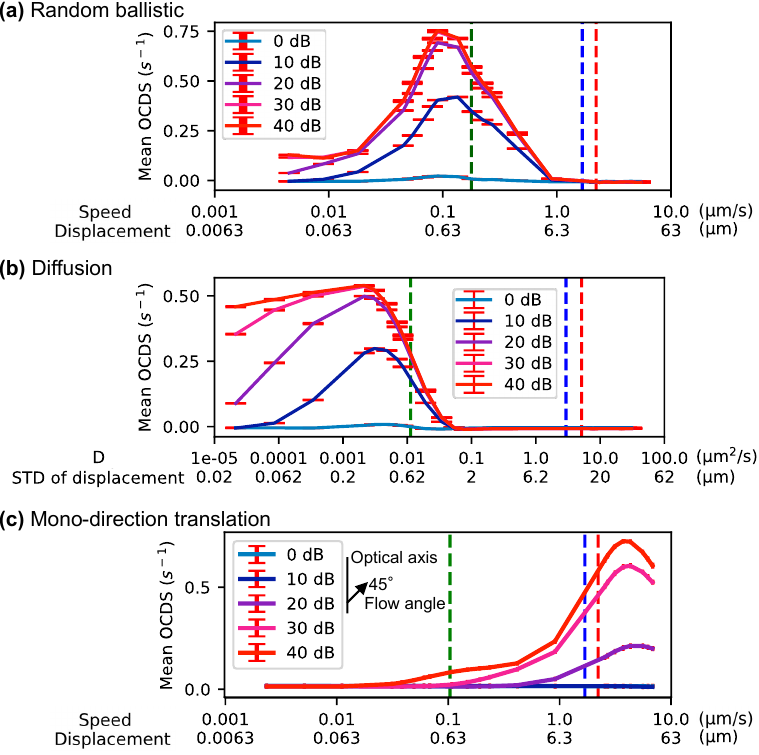}
	\caption{%
		Mean OCDS obtained via simulations.
		The results are presented in the same manner that was used in Fig.\@ \ref{fig:LIVNoise}.
		The OCDS was found to exhibit peaks at specific motion parameter values, and each peak becomes broader as the SNR decreases.
	}
	\label{fig:OCDSNoise}
\end{figure}
Figure \ref{fig:OCDSNoise} summarizes the OCDS curves obtained as functions of the motion parameters.
The curves show that the OCDS is sensitive to a specific range of the motion parameters, i.e., the OCDS curves show peaks around a specific motion parameter value.
These findings are consistent with the previously reported interpretations of OCDS signals obtained from cancer spheroids \cite{AbdElSadek2020,ElSadek2021}.
These OCDS peaks become broader as the SNR decreases.
In other words, the OCDS becomes less specific to a particular motion speed or diffusion coefficient in cases of lower SNR.

\begin{figure}
	\centering\includegraphics{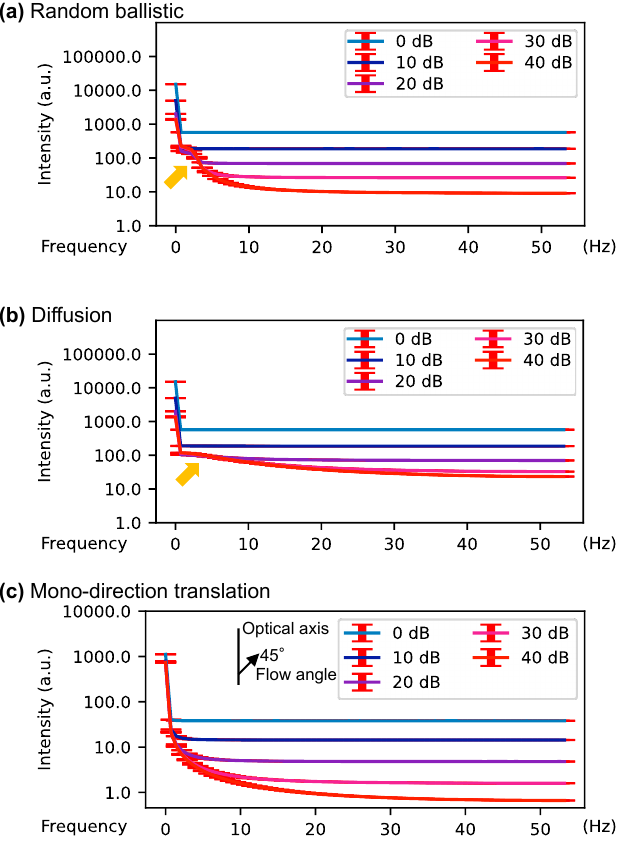}
	\caption{%
		Mean amplitude spectra obtained via the simulations.
		The results for the random-ballistic and diffusion motions show an elevation of the amplitude spectra at a particular frequency (indicated by yellow arrows).
		The error bars show the standard deviations for the simulation trials, but the values are very small.
		Additionally, this elevation becomes difficult to recognize as the SNR worsens.
	}
	\label{fig:APSNoise}
\end{figure}
Figure \ref{fig:APSNoise} summarizes the amplitude spectra acquired with several different SNRs.
In this case, we used a specific motion speed of 0.9 \um /s for the random-ballistic motion and the mono-directional translation, and a specific diffusion coefficient of 0.19 $\mu$m$^2$/s for the diffusion.
With the random-ballistic and diffusion motions, the characteristic elevation of the amplitude spectra (indicated by the yellow arrows), which represents the fingerprint of the specific motion parameter values, can be seen with the high SNR configurations, whereas it is not evident for the mono-directional flow.
As the SNR decreases, the noise backgrounds of the spectra increase, and the characteristic elevations become unrecognizable, even in the random-ballistic and diffusion motion examples.

\begin{figure}
	\centering\includegraphics{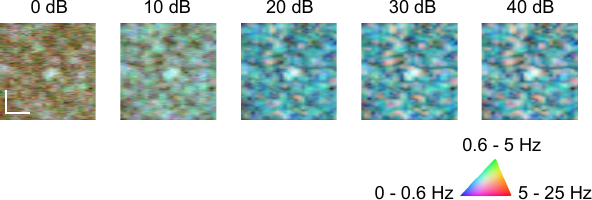}
	\caption{%
		Pseudo-color images of AS-DOCT acquired at several SNRs.
		The scatterers in this case are moving under the random-ballistic-motion model conditions.
		Although low to middle frequencies (light blue) are observed with the high SNR images, the frequency becomes  erroneously high (yellow) if the SNR is low.
		The scale bar represents 50 \um.
	}
	\label{fig:APSNoiseColor}
\end{figure}
To understand the noise effects in the AS-DOCT image intuitively, we generated pseudo-color AS-DOCT images of the AS-DOCT as shown in Fig.\@ \ref{fig:APSNoiseColor}. 
The motion here was random-ballistic motion with a scatterer speed of 0.9 \um /s, and the three spectral sub-bands are [0 Hz, 0.6 Hz] (blue), [0.6 Hz, 5 Hz] (green), and [5 Hz, 25 Hz] (red).
Although the high SNR case shows a light blue appearance, indicating low to middle frequencies, it erroneously rises to a high frequency (yellow) as the SNR decreases.

\subsection{Dynamic-scatterer-ratio (DSR) study}
\label{sec:dsrStudy}
As we mentioned in Section \ref{sec:3DsimualtionFlow}, our simulation framework can simulate cases in which the static and dynamic scatterers are mixed.
The mixing ratio is represented by the DSR, which is defined as the ratio of the number of dynamic scatterers to the total number of scatterers.

\subsubsection{Simulation and study protocol of the DSR study}
To compute the many cases with different DSRs effectively, we took the following steps.
First, we generated $M$ independent time sequences composed of complex OCT volumes with the same simulation configurations using the 3D-OCT-volume formation model.
All the scatterers are dynamic here, and thus this case is similar to the fully dynamic case, except that we did not add noise.
We refer to these volumes as dynamic volumes here.

Second, we generated another $M$ time sequences of complex OCT volumes in which the scatterers are static.
These OCT volumes are referred to as static volumes.
Because we also assumed that the noise does not exist, all OCT volumes within the same time sequence are perfectly identical.
Therefore, we generated only one OCT volume per sequence in practice.
Note here that the scatterer density for each static and dynamic volume was set at 1/$M$ of the target scatterer density of the study.

The final OCT sequence is generated from these volumes.
Specifically, at each time point, $m$ dynamic volumes and $M-m$ static volumes are combined (i.e., added together).
The intensity OCT volume, i.e., the 3D speckle pattern, is obtained as the absolute square of the complex volume obtained.
The DSR of the final volume is given by $m/M$.
By combining the volumes in different ways, we can generate 3D speckle patterns with different DSRs.
Even if the same DSR is used, we can also generate different realizations of the speckle by selecting different dynamic and static volumes.
This strategy greatly reduces the computation costs for OCT-volume-sequence generation.

In this study, the LIV was computed using the three motion types.
The number of dynamic sub-volumes $M$ = 10 and the DSR values were 0.1, 0.2, 0.3, $\cdots$, 0.9.
The DSR values of 0.0 and 1.0 were omitted here because the former is the fully static case and the latter was considered during the fully dynamic study (Section \ref{sec:fullyDyanmicStd}).
The scatterer density of the final volume was 0.055 scatterers/\cbum, i.e., the density of each sub-volume was one-tenth of that of the final volume. 
The SNR was considered to be infinite here (i.e., no noise was present).
The other parameters were the same as those used in the fully dynamic study.

\subsubsection{Results of the DSR study}
\begin{figure}
\centering\includegraphics{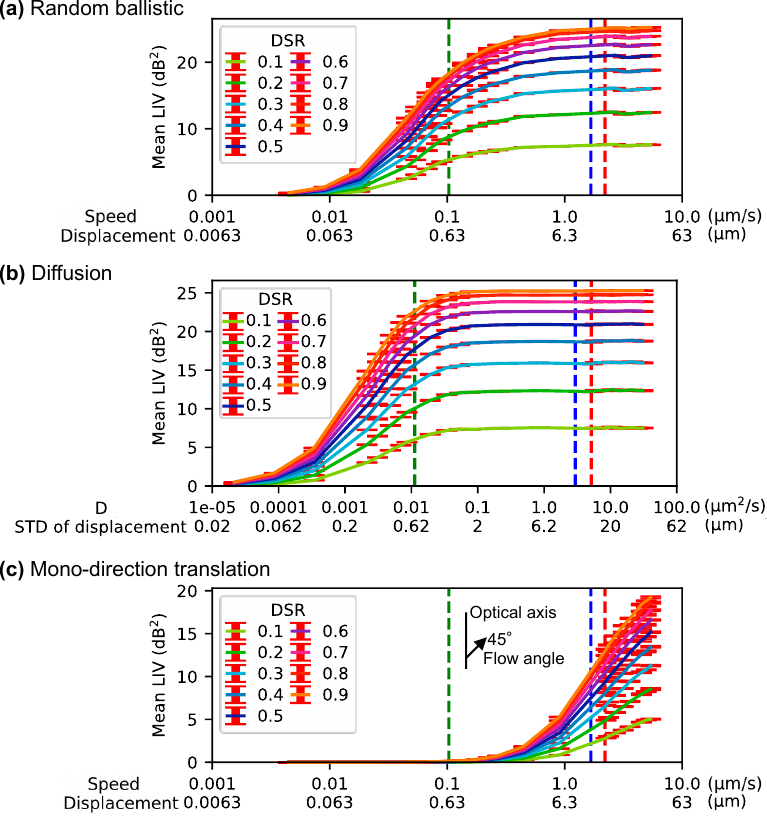}
	\caption{%
		DSR dependence of the LIV for the three motion types: (a) random ballistic motion, (b) diffusion, and (c) mono-directional translation. 
		The results are presented in the same manner as Fig.\@ \ref{fig:LIVNoise}.
		A larger LIV was observed with larger DSR values.
	}
	\label{fig:LIVDSR}
\end{figure}
In Fig.\@ \ref{fig:LIVDSR}, the results show that a higher DSR produces a higher LIV value for all motion types.
This indicates that the LIV is sensitive to the DSR.
This conclusion overrides our previous understanding of the LIV \cite{AbdElSadek2020}, in which the LIV was considered to be sensitive to the magnitude of the dynamics.
Additionally, we also noted that both the magnitude and the dynamics were not rigidly defined in this context, and thus the previous interpretation of LIV was ill defined.

\subsection{Examples of the single-point formation model}
Examples of the single-point formation model have been omitted from this paper, because the results of extensive studies using this model have been presented elsewhere.
For example, Section 3 in Ref.\@ \cite{Morishita2024a} provides comprehensive analyses of the LIV, OCDS, and two more new DOCT contrasts.
In another example, the wavelength- and resolution-dependences of both the LIV and the OCDS were investigated using the single-point formation model and compared with experimental results \cite{Fujimura2024, Fujimura2025arXiv}.

\subsection{Discussion of example studies}
\subsubsection{Motion-type dependence of LIV sensitivity}
For the random motions, including the random-ballistic and diffusion motion types, the LIV was found to begin to increase at a displacement of approximately 0.06 \um, which is approximately one-tenth of the wavelength for the double-pass configuration [Fig.\@ \ref{fig:LIVNoise}(a) and (b) for the fully dynamic study, and Fig.\@ \ref{fig:LIVDSR}(a) and (b) for the DSR study].
In contrast, for the mono-directional translation case, the elevation started at much larger displacement values (approximately ten times that in the random motion cases).

This behavior can be explained by the practical phase sensitivity of the LIV.
Because the LIV is computed from the logarithmic OCT intensity, it is thus not sensitive to the phase of the OCT signal.
However, in the random motion cases, the relative depth positions of the scatterers are altered by the motion, and these changes lead to mutual phase changes in the electric-field contributions  from each scatterer. 
This phase alternation results in either constructive or destructive interference, and alternates (scintillates) the intensity speckle pattern.
Such speckle scintillation can be observed in the simulated speckle patterns in Fig.\@ \ref{fig:speckleTimeSequence}(d, e) and the related supplementary videos (Visualizations 1 and 2).
The LIV can capture this speckle scintillation.
Because the mutual phase alternation originates from the sub-wavelength displacements of the scatterers, the LIV is then sensitive to the sub-wavelength displacements of the scatterers.

In contrast to the random motions, in the mono-directional translation motion, all scatterers travel at the same speed and in the same direction, and this maintains their relative distances.
In this scenario, the mutual phases do not alter, and thus the speckle pattern does not alter, but only shifts as preserving its shape.
This can be confirmed by the speckle pattern example shown in Fig.\@ \ref{fig:speckleTimeSequence}(f) and the related supplementary video (Visualization 3).
Because the speckle size is approximately the same as the optical resolution,  the LIV becomes sensitive only to larger motions on the scale of the resolutions (e.g., of a few micrometers).

In conclusion, our example studies and the speculations in this section indicate that the LIV is sensitive to sub-wavelength random motions.
However, if the motion is the mono-directional translation type, the LIV is sensitive only to resolution-scale motions.

\section{Discussion}  
\subsection{Voxel size selection}
It should be noted here that the voxel size of the 3D-OCT-volume formation model must be small enough when compared with the overall travel distance of the scatterers, otherwise, the speckle fluctuations cannot be simulated correctly. 

\begin{figure}
	\centering\includegraphics{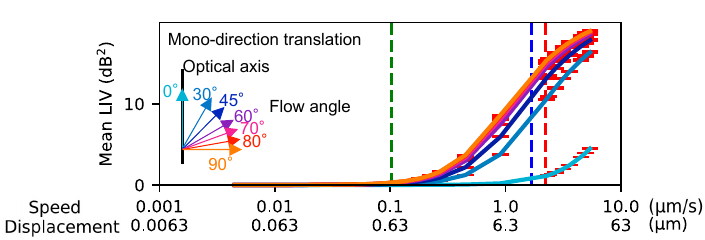}
	\caption{%
		Mean LIVs obtained from mono-directionally translating scatterers with several different translation (flow) angles.
		The results are presented in the same manner as the results in Fig.\@ \ref{fig:LIVNoise}.
		The zero-degrees case exhibits erroneously low LIV values because of the large axial size of the voxels.
	}
	\label{fig:LIVnoNoise}
\end{figure}
Figure \ref{fig:LIVnoNoise} demonstrates this effect.
The LIV here was computed for mono-directional motions at several flow angles.
The SNR was assumed to be infinitely high (i.e., no noise was present).
The voxel sizes of the OCT field were 1.95 \um (lateral) and 7.24 \um (axial), i.e., the voxel was longer in the axial direction.
In all cases other than the zero-degree flow angle case (i.e., where the flow is along the axial direction), the LIV becomes slightly smaller as the flow direction becomes more axial than lateral.
This small angle dependence may occur because the lateral resolution is smaller than the axial resolutions (FWHM values of 10.6 \um and 14 \um, respectively).
When compared with this moderate angle dependence, the zero-degree case shows a significant drop-off in the LIV values.
In this case, the scatterers only travel in the axial direction.
Because the axial size of the voxel is approximately four times larger than the lateral size, these scatterers are more likely to reside within the same voxel during the entire simulation time than to travel in the lateral directions.
Because the motions of each scatterer that are smaller than the voxel size cannot be captured by the 3D-OCT-volume formation model, this causes the artifactual reduction in the LIV. 

It was also noted that this artifact was particularly significant for the mono-directional motion for the following reasons.
Even when the displacements of the scatterers are smaller than the voxel size, the axial motion alters the phase of the complex reflectivity of each voxel (see the third paragraph of Section \ref{sec:3DsimualtionFlow}).
Additionally, this phase is sensitive to sub-wavelength motions.
For the random-ballistic and diffusion motion cases, the mutual phases of the scattered light beams are practically altered randomly by this effect, and the speckle pattern is greatly altered by even very small axial motions (see the movies corresponding to Fig.\@ \ref{fig:speckleTimeSequence}).
In contrast, for the mono-directional translation, all scatterers travel coherently and the distances between all scatterers remain constant.
As a result, the mutual phases also remain constant, and the speckle shape does not change.
Therefore, the artifact mentioned above becomes effectual.

Note here that in our example studies (Sections \ref{sec:fullyDyanmicStd} and \ref{sec:dsrStudy}), the voxel size is large in the axial direction.
However, the artifact is not effectual for the random-ballistic and diffusion motions, as we have discussed earlier, and it also may not be significant for the flow angle of 45 degrees because the scatterers have significant lateral velocity components.

\subsection{Future applications and directions}
Our simulation framework may have two major application phases.
Phase 1 is analysis of the imaging properties of DOCT, as demonstrated by the example studies (Section \ref{sec:exampleStudy}).
Similarly, this framework will be applicable to the analysis of several other types of OCT-based tissue dynamics imaging, such as OCT angiography (OCTA), which will be discussed in a later paragraph.

In Phase 2, the simulated results will be used to design better DOCT algorithms.
For example, Morishita \etal have used this simulation framework to design new DOCT algorithms that represent the DSR and the scatterer speed more directly than the LIV and the OCDS \cite{Morishita2024a}.
Furthermore, Fujimura \etal constructed an estimation theory to estimate the DSR and the scatterer speed quantitatively from the DOCT values \cite{Fujimura2025}.
Their estimator is a regression model that was made using the pairs of the set parameters from the DOCT simulation (the DSR and the scatterer speed) and the corresponding simulation results (i.e., the DOCT values).
In addition, the noise-effect simulation feature of the simulator will contribute to design DOCT algorithms with high noise immunity.

There are several features and improved functions that could be possibly included in the simulation framework in the future.
One possible improvement involves the OCT signal formation methods. 
The current framework uses a relatively simple OCT imaging model.
Although this model is fast and adequate, there are several image properties that cannot be recapitulated by this image model, such as absorption and depth-oriented signal attenuation.
This limitation can be resolved by introducing more sophisticated imaging models \cite{Schmitt1997,Thrane2000, Munro2015OpEx, Munro2016OpEx, Brenner2016, Macdonald2021BOE, Mao2024}.

Another feasible and relatively straightforward extension involves introducing more DOCT algorithms. 
In addition to the DOCT algorithms, OCTA algorithms, such as the amplitude-decorrelation angiography (SSADA) \cite{Jia2012}, optical microangiography (OMAG) \cite{Wang2010}, and phase-sensitive correlation-based OCA \cite{Makita2016} algorithms can be incorporated.
This incorporation would then enable numerical analysis of the imaging properties of OCTA.

The introduction of a bulk motion model is another important potential extension.
For example, \invivo samples frequently exhibit large but comparatively slow movements.
These bulk motions can be periodic or nonperiodic, can be linear or nonlinear, and can incorporate rotations rather than include only shifts.
By introducing a new software module to emulate such motion and using it along with the intracellular/intratissue motion models, we can analyze the imaging properties of both \invivo DOCT and OCTA.

\section{Conclusion}
In this paper, we have proposed a DOCT simulation framework that comprises three types of intracellular/intratissue motion models, OCT signal formation models that are incorporated with physically realistic noise models, and DOCT algorithms.
The motion models were designed based on seven intracellular/intratissue activity types.
For OCT signal formation, two models were implemented, including the less-computationally-intensive single-point formation model and the 3D-OCT volume formation model, which enables volumetric speckle generation.
The DOCT algorithms include only the LIV, OCDS, and amplitude spectrum algorithms, but it is easy to include additional algorithms.
The utility of the proposed simulation framework was demonstrated via several example studies.

The framework presented here will give deeper understanding of the imaging properties of DOCT and will serve for further improvements in DOCT methods.
Since this framework is available as open-source \cite{GitHubDoctSim}, it will greatly contribute to the technology and application communities of DOCT.

\section*{Funding}
Core Research for Evolutional Science and Technology (JPMJCR2105); 
Japan Society for the Promotion of Science (21H01836, 22F22355, 22KF0058, 22K04962, 24KJ0510);
Japan Science and Technology Agency (JPMJFS2106). 

\section*{Disclosures}
Feng, Fujimura, Lim, Seesan, Morishita, El-Sadek, Mukherjee, Makita, Yasuno: Sky Technology(F), Nikon(F), Kao Corp.(F), Topcon(F), Panasonic(F), Santec (F), Nidek (F).
Feng, Fujimura, and Lim have been appointed to positions at Vivolight, Hitachi High-Tech Science Corp., and Ainnovi, respectively.

\section*{Data availability}
The data that support the findings of this study are available from the corresponding author upon reasonable request.
The open-source implementation of the presented simulation framework is available at an online repository \url{https://github.com/ComputationalOpticsGroup/COG-dynamic-OCT-contrast-simulation-library} \cite{GitHubDoctSim}. 

\bibliography{DOCTsimulator}

\end{document}